# Application of Different Metaheuristic Techniques for Finding Optimal Test Order during Integration Testing of Object Oriented Systems and their Comparative Study

Chayanika Sharma*[1], Ritu Sibal[2]
[1,] Netaji Subhas Institute of Technology, University of Delhi, New Delhi, India
[2,] Netaji Subhas Institute of Technology, University of Delhi, New Delhi, India
Email: [1]chayanika_29a@yahoo.com, [2]ritusib@hotmail.com

***Abstract.*** **In recent past, a number of researchers have proposed genetic algorithm (GA) based strategies for finding optimal test order while minimizing the stub complexity during integration testing. Even though, metaheuristic algorithms have a wide variety of use in various medium to large size optimization problems [21], their application to solve the class integration test order (CITO) problem [12] has not been investigated. In this research paper, we propose to find a solution to CITO problem by the use of a GA based approach. We have proposed a class dependency graph (CDG) to model dependencies namely, association, aggregation, composition and inheritance between classes of unified modeling language (UML) class diagram. In our approach, weights are assigned to the edges connecting nodes of CDG and then these weights are used to model the cost of stubbing. Finally, we compare and discuss the empirical results of applying our approach with existing graph based and metaheuristic techniques to the CITO problem and highlight the relative merits and demerits of the various techniques.**

* Corresponding Author:
Chayanika Sharma,
Faculty of Computer Science and Engineering,
Delhi University, India,
Email: chayanika_29a@yahoo.com   Tel: +91 9911718232

## 1. Introduction

Software testing represents a substantial percentage of the cost of developing software systems. Typical percentages range between 30% and 60% [1]. Software testing is carried out to detect presence of faults and to evaluate whether the software works correctly as specified by requirements. However, software testing is a time consuming and expensive task [2], [3]. Testing of OO software is different and difficult than traditional software, considering the complex dependencies that exist between classes due to generalization and client-server relationships. One of the challenges in testing of OO systems is to perform integration testing of classes containing thousands of circuits, potentially denoting as many ways for class instances to interact[4] [5].

Integration testing is a software testing technique that aims to find errors associated with the interfacing and integration of system components [6]. In OO testing, it is difficult to determine the order in which classes are to be integrated and tested [7]. One of the main difficulties for cost-efficient integration of components is minimization of the number of stubs to be written. Stub is a place holder for a class that does not implement the full functionality but only necessary partial functionality for its compilation and integration [8]. A stub is created for every removed class in the system, thus increasing cost of integration. Stubs can be modeled as: specific stubs and realistic stubs. A specific stub simulates the services for use of a given client only. Realistic stub simulates all services that the original class can provide [9]. In CITO problem, the goal is to determine the order in which classes can be integrated so as to minimize stub construction cost.

The CITO problem is very important to industry, most companies solve it very badly, leading to wasted resources and faulty software. Moreover, the research community has published several good solutions that are either not in use or very costly. Most companies integrate groups of classes that are big enough so that there are no cycles. This makes testing difficult, debugging much more expensive, and ultimately allows more integration faults to slip through into system testing. Several methodologies and strategies have been proposed for finding the optimal test order during integration testing of OO systems and solving CITO problem. These strategies are primarily of two types [7] viz. GA based strategies and graph based strategies. GA is an evolutionary algorithm based on the concepts from biological evolution [3] [10]. In graph based approach, the system space is represented as a directed graph. Graph based testing strategies are unable to deal with breaking dependencies among components, thereby making it difficult to identify the integration test order [13]. In graph based approach, the dependency cycle sorting is applied to remove the cyclic dependency between components. Integration test order is then obtained by applying reverse topological sorting. However, in real life problems the topological sorting is not just difficult but impossible [12].

The major issue in integrating OO software is with cycles. When breaking a cycle, we must create "stubs" to provide partial implementations of classes that have not yet been integrated. The goal is to break cycles in such a way that the cost of these stubs is minimized. This paper suggests a new genetic algorithm based approach for computing the order in which classes should be integrated.

This paper is divided into eight sections. In section 2, we present an overview of related work. Section 3 describes the modeling of UML class dependencies using CDG. Section 4 describes in detail our proposed approach. Section 5 presents experimental results of our proposed approach. In Section 6, we compare the results of our approach with other existing graph based and GA based approach. Section 7 discusses the empirical results of applying GA, micro-GA and cuckoo search algorithm on CITO problem. Finally, section 8 concludes our research work.

## 2. Related Work

Thierry Jeron et al. [11] have proposed a graph model called test dependency graph (TDG) for the representation of test dependency in OO systems. The approach focuses on three types of dependencies: 'method to method', 'class to class' and 'method to class' to find the optimal test order. Jutarat et al. have proposed an approach to reduce the number of stubs in a test order using OO slicing technique and TDG. Class slicing is a kind of partial testing in OO slicing technique. The strategy uses Tarjan's algorithm [14] for finding strong connected component (SCC). SCC of a graph is a sub graph such that for any pair of vertices u, v, there is a path from u to v and vice versa. Two vertices are equivalent if they are in the same SCC [11]. Tai and Daniel [15] have proposed a two stage algorithm to remove dependency cycles. Le Traon et al. [16] have proposed an alternative strategy based on graph search algorithms that identifies SCC and yields more optimal results. Kung et al. [17] point out that association relationships are usually the weakest links in a class diagram and stated that every cycle in a class diagram contains at least one association relationship. C. Briand et. al. [4] have systematically reviewed and analyzed existing graph based techniques to generate test orders having minimum number of stubs. The existing techniques for generating test orders while integrating classes by Briand et.al. , Le Traon and Tai and Daniels have been investigated on different case studies. Aynur Abduzarik et. al. [12] used new technique and proposed a graph based algorithm to solve CITO problem. The test dependencies between classes were modeled using weighted object relation diagram (WORD). The weights to WORD are assigned to represent cost of creating stubs. The algorithm uses edge weights and node weights to solve CITO problem.

In recent past, other approaches to find optimal test order have also been proposed [13] [18]. These are not graph based but make use of GA. GA has emerged as a practical, robust optimization technique and search method. A GA is a search algorithm that is inspired by the way nature evolves species using natural selection of the fittest individuals. Briand [18] has used GA to find the optimal test order by measuring stub complexity. The composition and inheritance relationships are considered as strong relationships that cannot be broken. The complexities of association, aggregation and usage dependencies were computed based on the level of coupling between client class and server class. Coupling is measured using attribute dependency and method dependency. The cost function is defined by attribute dependency and method dependency to be minimized by GA. Vu le Hanh et. al. [9] used GA to find the optimal integration order by using number of stubs as a cost function to be minimized. Triskell's strategy has been adopted in their work.

Erik Arisholm et. al. [20] calculates percentage of intensity of interaction in dynamic model by determining the messages sent between objects in the UML sequence diagram. In our approach, we adopt Arisholm's strategy by determining number of methods and attributes shared or inherited among classes by aggregation, composition, association and inheritance test dependencies to determine coupling strength (CS). Cycle weight (CW) to determine weight among classes is adopted from Aynur Abduzarik's strategy [12]. Number of cycles in CDG determines the CW for each edge. In our work, stub complexity is determined by three parameters namely CS, CW and information flow (IF) metric [27]. IF determines the information flow complexity associated with each class or component in a class diagram. Also, we have proposed CDG to show test dependencies in the OO system. Thereafter, GA is applied to find the test order having stub of minimum complexity and thereby minimizing cost during integration testing of OO system. Although GA has already been used by researchers to solve CITO problem, in our research we have used a different strategy to find the cost of each test stub and also applied cuckoo search algorithm and Micro-GA to solve CITO problem.

Micro-GA and cuckoo search algorithm has not been applied so far for determining optimal test order during integration testing of OO system. Micro-GA refers to small population size with reinitialization [25]. Micro GA was first implemented by K. Krishankumar [26], with population size of 5, crossover rate of 1 and mutation rate of 0.0. Only best candidates are carried to the next generation. Tournament selection strategy is used for selection of individuals. Cuckoo search algorithm is proposed by X. S Yang and S. Deb in 2010. In cuckoo search algorithm, each generation is represented by a set of host nests each carrying an egg or the solution. The number of solutions remains fixed in each generation. Cuckoo bird lays her egg in the nest of another host species. The host bird discovers egg of another species with a probability $p_a \epsilon$ [0, 1]. On discovering egg of another species, the host bird either throws the egg or destroys the nest. The new egg is formed by modifying one solution randomly. If new solution is better than existing solution, replace existing solution. In cuckoo search, elitism is adopted where only best individuals are carried over to the next generation. Worst solutions of fraction $p_a$ are discarded when new solutions are built. Cuckoo search algorithm has been applied to solve design optimization problem [21] [22] [23]. In X. S. Yang et. al. work [21], cuckoo search algorithm is simulated and compared with GA and particle swam optimization. The results of cuckoo search algorithm have been found superior for multimodal objective functions than GA and particle swam optimization. Micro-GA has been applied for solving optimization problems [24] [25]. Y. Quin et. al. [24], applied Micro-GA to optimize time domain ultra wide antenna array. In K. KrishanKumar [26] work, results of Micro-GA are better and faster when compared with simple GA (population size = 50, crossover rate = 0.6 and mutation rate = 0.001) on two stationary functions and real world control engineering problem.

## 3. Modeling Class to Class Dependency

Several relationships are used to model dependencies in static and dynamic UML diagrams. Some of relationships used to model dependencies are association, aggregation, composition and inheritance. In this section, we define CDG to model various class to class dependencies in static UML class diagram**.**

## 3.1 UML TO CDG Mapping

The association, aggregation, composition and inheritance relationships are commonly used to exhibit dependencies between classes of UML class diagram. Figure 1 shows the rules to map aggregation, composition, association and inheritance relationship in UML class diagram into CDG**.**

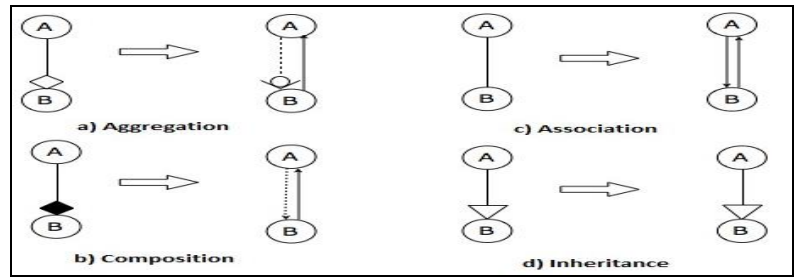

**Figure 1.** CDG Mappings

The mapping between component A and component B of a CDG are summarized below:-
*a) Aggregation*: - We use the term optional dependency in our work and this is represented as a dotted arrow with 'o' symbol on the top. The aggregation is mapped into association relationship from B to A and optional dependency from A to B. The optional dependency from A to B implies that aggregate B is optional for A because in aggregation, the aggregate part is not mandatory for part component and the owner may change over time.
*b) Composition*: - In composition relationship, the lifetime of component A is dependent on aggregate component B. Therefore, composition relationship is shown as the dependency between A on aggregate B and association from B to A.
*c) Association*: - The association relationship between component A and component B is shown as bidirectional association from A to B and B to A.
*d) Inheritance*: - The inheritance relationship between component A and component B remains the same and is left untouched.

## 4. Proposed Approach
In the present research work, our aim is to identify the integration test order for an OO system that results into minimum cost of stub construction. To do so, we have applied GA based approach. GA is useful for making decent approximations with very large data sets. In our paper, CITO problem is applied on large sized case study. GAs are useful when the data sets get large, whereas detailed analysis algorithms almost always yield better approximations but blow up with large data sets. A practical disadvantage of GA involves longer running times on the computer.

To model the dependencies namely, association, aggregation, composition and inheritance between classes of UML model we propose a CDG. A weight is assigned to each edge of a CDG. Higher is the weight of an edge, higher is the stub complexity of the stub required to break a dependency between classes of CDG. The procedure for assigning weight to an edge of a CDG is illustrated in section 5.In our GA based approach, a weight or fitness value is assigned to each edge of CDG. Initially, a UML class diagram is converted into a CDG, where the nodes represent classes and edges represent the structural dependencies between them [6] [11]. In CDG, the association, aggregation and composition relationships of a class diagram are mapped into bidirectional edges. The weight of each edge connecting client and server class in the CDG is determined by CS, CW and IF. We define these three parameters as follows:-
*Coupling strength (CS)*:- A class is coupled with another class, if methods or instance variables of one class use methods or attributes of the other [4]. In our work, CS between server class and client class is determined by the number of methods and attributes of the server class used by client class having aggregation, composition, association or inheritance dependency divided by the number of methods and attributes in server and client class as shown in equation 1.
*CS = Number of methods and attributes used by client class / Total number of methods and attributes in server and client class*             (1)

In composition dependency, there exist a whole – part relationship where life time of attributes or members of a whole classifier is dependent on part classifier. If a whole classifier is deleted, then part classifier also gets deleted. Although the part classifier can be removed before whole classifier is deleted. In inheritance relationship, if the parent class or superclass is deleted then all the subclasses or child classes also get deleted. Therefore, the removal of inheritance and composition dependency will lead to deletion of subclasses and part classifier. In our work, we have considered composition and inheritance as strongest relationship as breaking them will lead to high stub complexity. Consequently in order to reduce stub construction cost we try to prevent breaking inheritance and composition dependencies.

*Cycle Weight (CW)*:- CW of an edge is determined by identifying number of cycles the edge is involved in divided by total number of cycles in the SCC.

*Information Flow (IF)*:- IF determines the information flow complexity associated with each class [27]. IF for each class A or IF (A) is calculated by applying equation given below:-

$$IF(A) = FANIN(A) * FANOUT(A) \qquad (2)$$

Where, Fan In is number of components calling A and Fan out is number of components called by A. To avoid large computational values, we have changed multiplicative operator in equation 2 to additive operator for determining IF (A) as shown by equation 3.

$$IF(A) = FANIN(A) + FANOUT(A) \qquad (3)$$

After converting a class diagram into a CDG, SCC is identified using Tarjan's algorithm. GA operators are applied to determine the test order having test stubs of minimum complexity. Finally, integration test order with minimum complexity for writing test stubs is obtained. The edges having minimum test dependencies will have least stub complexity, if broken. In other words, breaking the dependency of an edge connecting two classes having high CS, CW and IF will lead to high stub complexity if broken. Higher the level of CS, CW and IF between client and server class, higher will be the cost of breaking dependencies between them. The procedure is outlined as steps below.

**Procedure**
Steps involved for identifying the optimal test order during class integration testing are outlined as following:-
1. Convert the UML class diagram into CDG $G_i = (V_i, E_i)$
2. Apply Tarjan's algorithm to identify the SCC in $G_i$.
3. Identify the cycles involved in the $SCC_i \in G_i$.
4. Assign weight $W_i$ to each edge $E_{scci} \in SCC$ by applying equation 4.

$$W_{scci} = k * \frac{CW * IF}{CS} \qquad (4)$$

Where, k is a constant and is 0 for inheritance and composition dependency and 1 for other dependencies. We assume, if the dependencies between client and server class is zero, then dependencies are not supposed to be broken. CS is the number of methods and attributes of a server class inherited or shared by a client class having association, aggregation, composition and inheritance test dependencies divided by total number of methods and attributes in server and client class. CW is the number of cycles an edge is involved in divided by total number of cycles in the SCC and IF is information flow complexity associated with each node or class determined using equation 3. Consequently, $W_{scci}$ is used as a fitness function in GA.

5. Selection: - Possible test orders in $SCC_i$ form the chromosome or the candidates for initial population. Therefore, in our case study a class represents a gene and a string of possible classes forming a test order represents a chromosome. The permutation encoding technique [28] is used to represent the solution to our problem. For a set of class {A, B, C, D, E and F} a possible chromosome sequence is {B, C, E, F, A, and D}.

6. Crossover: - There are number of techniques of crossover, but all require swapping of genes or sequence of bits in the chromosome. It involves swapping between two individuals or test data in our case. One point crossover is applied in our work. In general, constant crossover and mutation probability is applied in GA. In Bo Zhang and Chen Wang work [19], adaptive crossover and mutation probability has been applied. The crossover probability $P_c$ and mutation probability $P_m$ is high for bad individuals having low fitness value and is low for good individuals having high fitness value. Based on the Bo Zhang's approach [19], we have assigned high crossover probability $Pc_1 = 75$ to the bad test orders having fitness value greater than average fitness value and lower crossover probability $Pc_2 = 20$ to good test orders having fitness value less than the average fitness value. This prevents good test orders or individuals from being modified.

7. Mutation: - Mutation is performed to introduce new traits or bring diversity in the population to avoid local optima. In mutation the bits are flipped from 0 to 1 and vice versa. Mutation probability $P_m$ is different for individuals having high and low fitness value. High mutation probability $Pm_1 = 25$ is assigned to bad individuals or bad test orders having fitness value higher than average fitness value and low mutation probability $Pm_2 = 15$ is assigned to good individuals or good test orders having fitness value less than average fitness value.

8. If all or set of edges have same weight, four rules have been proposed for converting cyclic CDG to an acyclic CDG, to be able to perform integration testing. These are as follows:-

>    RULE 1. Remove an edge involved in a cycle having weight $W_i$ higher than other edges in the SCC.
>    RULE 2. Remove the optional test dependency (OTD) or use dependency relationship in case of tie between association test dependencies.
>    RULE 3. In case of tie between two association test dependencies, edge having higher weight $W_i$ is removed first.
>    RULE 4. In case, two associations under test have same weights then delete the edges traversed first in depth first search (DFS) call of Tarjan's algorithm.

We have assumed use dependency relationship as the weakest relationship among association, aggregation, composition and inheritance relationships. A dependency relationship exists when one class requires another class for its execution [11].

```
1.   Convert class diagram into CDG, G_i = (V_i, E_i).
2.   Apply Tarjan's algorithm to identify SCC_i ∈ G_i.
3.     If SCC exists in a G_i
4.        for each ( SCC_i(V_scci, E_scci) ) do
5.           find total number of cycles in SCC_i
6.             for each E_SCCi do
7.                Remove E_SCCi in decreasing order by weight W_i determined using equation 4.
8.                If E_scci have equal W_i and CW_i
9.                   Remove E_scci having dependency or optional test dependency
10.                  Else remove edge traversed first in DFS call of Tarjan's algorithm.
11.            end for
12.         Generate test order T_i ∈ SCC_i or initial population randomly.
13.           for each ( T_i ∈ SCC_i ) do
14.              Determine fitness value or W_i.
15.              Apply crossover and mutation operators to generate individuals in the new population.
16.              If W_i > Fitness_Average , P_c1 = 75% and P_m1 = 25%
17.                Else P_c2 = 25%  and P_m2 = 15%
18.              End If
19.           Repeat GA process until termination criteria is reached.
20.        End for
21.     End for
22.   End if
```

Figure 2. Algorithm for the Proposed Approach

9. Apply Topological sorting algorithm to construct integration test order for the graph $G_i$.
Algorithm of our proposed approach is shown in Figure 2.

## 5. Case Study

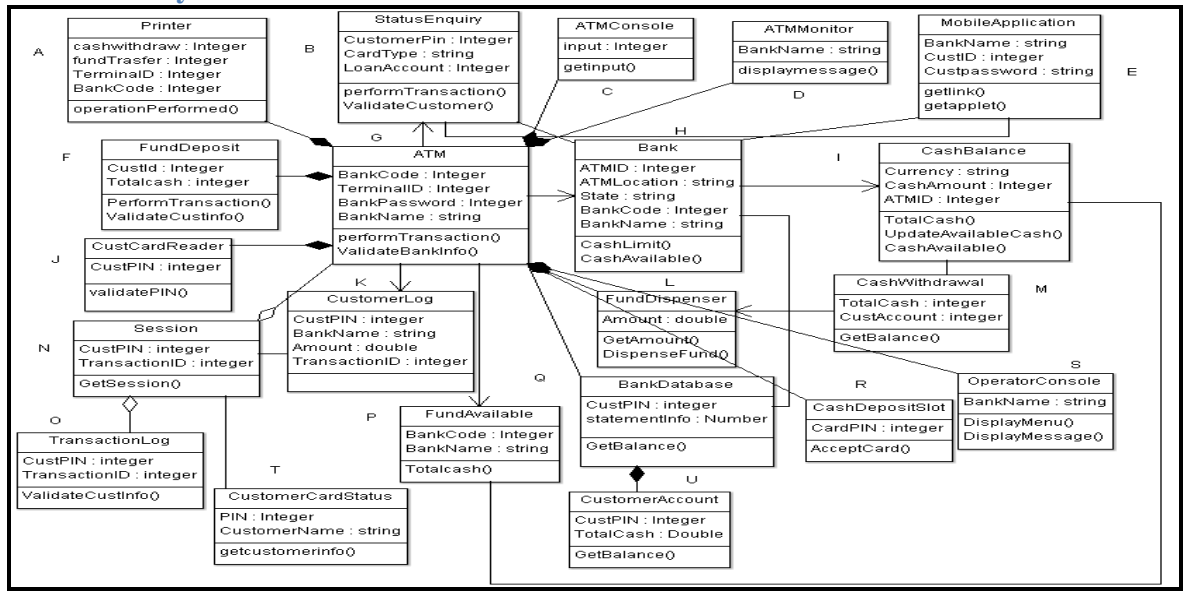

Figure 3. ATM Class Diagram

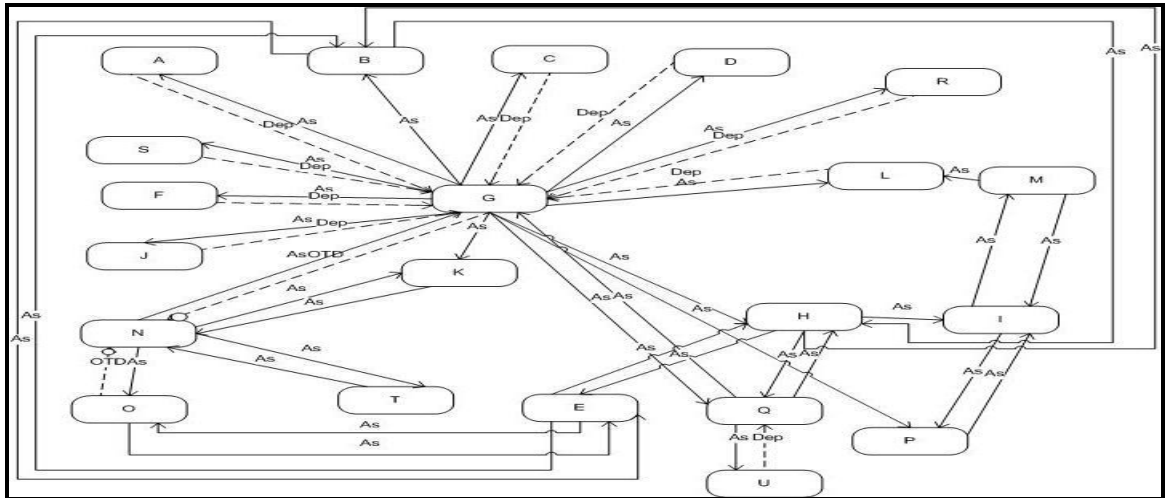

**Figure 4.** CDG of ATM Class Diagram (As -> Association, Dep -> Use Dependency, OTD -> Optional Test Dependency)

The proposed approach is applied on the class diagram of an ATM system shown in Figure 3. The association, aggregation and composition relationships are used to show the dependencies between classes. The class diagram of Figure 3 is converted into a CDG as shown in Figure 4. The modeling of various test dependencies while converting ATM class diagram into CDG is shown in Table 1.

**Table 1.** ATM Class Diagram to CDG Modeling

| S.NO | RELATIONSHIP | CLASSES | PROPERTIES | CDG MODELING |
|---|---|---|---|---|
| 1. | Composition | Printer and ATM ( A and G) | whole classifier (G) part classifier (A) | A->G (Use Dependency) G->A (Association) |
|  |  | Fund Deposit and ATM ( F and G) | Whole classifier (G) Part classifier (F) | F->G (Use Dependency) G->F (Association) |
|  |  | Customer Card Reader and ATM (J and G) | Whole classifier (G) Part classifier (J) | J->G (Use Dependency) G->J (Association) |
|  |  | ATM Console and ATM (C and G) | Whole classifier (G) Part classifier (C) | C->G (Use Dependency) G->C (Association) |
|  |  | ATM Monitor and ATM (D and G) | Whole classifier (G) Part classifier (D) | D->G (Use Dependency) G->D (Association) |
|  |  | Operator Console and ATM (S and G) | Whole classifier (G) Part classifier (S) | S->G (Use Dependency) G->S (Association) |
|  |  | Cash Deposit Slot and ATM (R and G) | Whole classifier (G) Part classifier (R) | R->G (Use Dependency) G->R (Association) |
|  |  | Fund Dispenser and ATM (L and G) | Whole classifier (G) Part classifier (L) | L->G (Use Dependency) G->L (Association) |
|  |  | Customer Account and Bank Database (U and Q) | Whole classifier (Q) Part classifier (U) | U->Q (Use Dependency) Q->U (Association) |
| 2. | Aggregation | ATM and Session ( G and N) | Whole Classifier (G) Part classifier (N) | N->G (Association) G->N (Optional Test Dependency) |
|  |  | Session and Transaction Log (O and N) | Whole classifier (N) Part classifier (O) | N->O (Association) O->N (Optional Test Dependency) |
| 3. | Association | ATM and Status Enquiry ( G and B) | Unidirectional Association | G->B (Association) |
|  |  | Status Enquiry and Mobile Application (B and E) | Bidirectional Association | B->E (Association) E->B (Association) |
|  |  | Bank and Mobile Application (H and E) | Bidirectional Association | H->E (Association) E->H (Association) |
|  |  | ATM and Bank (G and H) | Unidirectional Association | G->H (Association) |
|  |  | Bank and Cash Balance ( H and I) | Unidirectional Association | H->I (Association) |
|  |  | Bank and Bank Database (H and Q) | Bidirectional Association | H->Q (Association) Q->H (Association) |
|  |  | ATM and Bank Database (G and Q) | Bidirectional Association | G->Q (Association) Q->G (Association) |
|  |  | Cash Balance and Fund Available ( I and P) | Bidirectional Association | I->P (Association) P->I (Association) |
|  |  | ATM and Fund Available (G and P) | Unidirectional Association | G->P (Association) |
|  |  | Cash Balance and Cash Withdrawal (I and M) | Bidirectional Association | I->M (Association) M->I (Association) |
|  |  | Cash Withdrawal and Fund dispenser (M and L) | Unidirectional Association | M->L (Association) |
|  |  | ATM and Customer Log (G and K) | Unidirectional Association) | G->K (Association) |
|  |  | Session and Customer Log (N and K) | Bidirectional Association | K->N (Association) N->K (Association) |
|  |  | Session and Customer Card Status (N and T) | Bidirectional Association | N->T (Association) |

| | | | | T->N (Association) |
|---|---|---|---|---|
| | | Mobile Application and Transaction Log (E and O) | Bidirectional Association | E->O (Association) |
| | | | | O->E (Association) |
| | | Status Enquiry and Bank (B and H) | Bidirectional Association) | B->H (Association) |
| | | | | H->B (Association) |

The SCC identified by applying Tarjan's algorithm in CDG of ATM is {1, 2, 3, 4, 5, 6, 7, 8, 9, 10, 11, 12, 13, 14, 15, 16, 17, 18, 19, 20 and 21}. The cycles identified in CDG are shown below:-

1. A -> G -> A
2. C -> G -> C
3. D -> G ->D
4. R -> G -> R
5. L -> G -> L
6. E -> H-> E
7. B -> H -> B
8. S -> G -> S
9. F -> G -> F
10. J->G->J
11. G->N->G
12. N->O->N
13. N->T->N
14. K->N->K
15. G->Q->G
16. G->H->I->P->G
17. G->H->Q->G
18. Q->U->Q
19. G->K->N->G
20. I->P->I
21. E->O->E
22. I->M->I
23. B->E->B
24. G->H->I->M->L->G
25. G->H->E->O->N->G
26. H->G->B->E->H
27. G->P->I->M->L->G
28. H->Q->H

Table 2. Edge Weights for SCC {1, 2, 3, 4, 5, 6, 7, 8, 9, 10, 11, 12, 13, 14, 15, 16, 17, 18, 19, 20, 21}

| S. No. | Edge | CS | CW | IF (A-> B) = IF (A) + IF (B) | $W_i$ / Fitness Function | S. No. | Edge | CS | CW | IF (A-> B) = IF (A) + IF (B) | $W_i$ / Fitness Function |
|---|---|---|---|---|---|---|---|---|---|---|---|
| 1. | A->G | 1/11 | 1/28 | 26 | 11.56 | 25. | 5->2 | 4/10 | 1/28 | 11 | 1.1 |
| 2. | G->A | 2/11 | 1/28 | 26 | 5.78 | 26. | 8->5 | 1/12 | 2/28 | 13 | 11.38 |
| 3. | F->G | 2/10 | 1/28 | 26 | 5.2 | 27. | 5->8 | 1/12 | 2/28 | 13 | 11.38 |
| 4. | G->F | 1/10 | 1/28 | 26 | 10.4 | 28. | 7->8 | 6/13 | 4/28 | 31 | 9.43 |
| 5. | J->G | 1/8 | 1/28 | 26 | 8.32 | 29. | 8->9 | 1/13 | 2/28 | 12 | 10.5 |
| 6. | G->J | 2/8 | 1/28 | 26 | 4.16 | 30. | 8->17 | 3/10 | 2/28 | 13 | 3.10 |
| 7. | C->G | 3/8 | 1/28 | 26 | 2.74 | 31. | 17->8 | 2/10 | 1/28 | 13 | 2.6 |
| 8. | G->C | 2/8 | 1/28 | 26 | 4.16 | 32. | 7->17 | 5/9 | 1/28 | 30 | 2.14 |
| 9. | D->G | 2/8 | 1/28 | 26 | 4.16 | 33. | 17->7 | 2/9 | 2/28 | 30 | 9.74 |
| 10. | G->D | 1/8 | 1/28 | 26 | 8 | 34. | 9->16 | 1/9 | 2/28 | 8 | 5.19 |
| 11. | S->G | 2/11 | 1/28 | 26 | 5.78 | 35. | 16->9 | 1/9 | 2/28 | 8 | 5.09 |
| 12. | G->S | 1/9 | 1/28 | 26 | 9.45 | 36. | 7->16 | 2/9 | 1/28 | 27 | 4.90 |
| 13. | R->G | 1/8 | 1/28 | 26 | 8.32 | 37. | 9->13 | 3/9 | 3/28 | 8 | 2.67 |
| 14. | G->R | 4/8 | 1/28 | 26 | 2.08 | 38. | 13->9 | 2/9 | 1/28 | 8 | 1.45 |
| 15. | L->G | 2/9 | 3/28 | 27 | 13.5 | 39. | 13->12 | 2/6 | 2/28 | 6 | 1.30 |
| 16. | G->L | 3/9 | 1/28 | 27 | 3.27 | 40. | 7->11 | 1/10 | 1/28 | 27 | 10.8 |
| 17. | U->Q | 2/6 | 1/28 | 8 | 0.97 | 41. | 11->14 | 2/7 | 2/28 | 11 | 2.66 |
| 18. | Q->U | 1/6 | 1/28 | 8 | 1.89 | 42. | 14->11 | 2/7 | 1/28 | 11 | 1.52 |
| 19. | N->G | 3/9 | 3/28 | 32 | 10.40 | 43. | 14->20 | 2/6 | 1/28 | 10 | 1.21 |
| 20. | G->N | 2/9 | 1/28 | 32 | 5.82 | 44. | 20->14 | 2/6 | 1/28 | 10 | 1.21 |
| 21. | N->O | 1/6 | 1/28 | 10 | 2.35 | 45. | 5->15 | 1/8 | 2/28 | 8 | 4.40 |
| 22. | O->N | 2/6 | 1/28 | 10 | 1.21 | 46. | 15->5 | 3/8 | 1/28 | 8 | 0.84 |
| 23. | G->B | 3/11 | 1/28 | 29 | 4.30 | 47. | 2->8 | 3/12 | 1/28 | 12 | 1.92 |
| 24. | B->E | 2/10 | 2/28 | 11 | 3.93 | 48. | 8->2 | 4/12 | 1/28 | 12 | 1.45 |

For space reasons, variables used between client and server classes are not shown in the ATM class diagram. Weight $W_i$ for each edge in a CDG is determined by applying equation 3. The edge A->G has CS A where server class G is using one variable from class 1. Hence its CS is 1/11. The CW for edge is determined by number of times edge is involved in cycle divided by total number of cycles in CDG. The edge A->G appears in cycle 1 and the total number of cycles in ATM CDG is 28, hence CW of edge A->G is 1/28. The IF is determined by adding the IF complexity of class A and class G by applying equation 3. The edge G->A has CS 2/11, as client class A is using data members Bank code and TerminalID from server class G. The edge G->A is involved in cycle 1; hence its CW is 1/28 and IF of edge G -> A is 24 + 2 = 26.

**Table 3.** Cyclic to Acyclic CDG of ATM Class Diagram

| S. No. | Edge Removed | Weight of edge removed | Cycles Removed | S. No. | Edge Removed | Weight of edge removed | Cycles Removed |
|---|---|---|---|---|---|---|---|
| 1. | L->G | 13.5 | {5, 24, 27} | | G->B | 4.30 | {26} |
| 2. | A->G | 11.56 | {1} | 16. | G->C | 4.16 | {2} |
| 3. | G->K | 10.8 | {19} | | D->G | 4.16 | {3} |
| 4. | H->E | 11.38 | {6, 25} | 17. | B->E | 3.93 | {23, 26} |
| 5. | E->H | 11.38 | {6, 26} | | G->L | 3.27 | {5} |
| 6. | H->I | 10.5 | {16, 24} | 18. | H->Q | 3.1 | {17, 28} |
| 7. | G->F | 10.4 | {9} | | C->G | 2.74 | {2} |
| 8. | N->G | 10.4 | {11, 19, 25} | 19. | I->M | 2.67 | {22, 24, 27} |
| 9. | Q->G | 9.74 | {15, 17} | 20. | K->N | 2.66 | {14, 19} |
| 10. | G->S | 9.45 | {8} | | Q->H | 2.60 | {28} |
| | G->H | 9.43 | {16, 17, 24, 25} | 21. | N->O | 2.35 | {12} |
| 11. | J->G | 8.32 | {10} | | G->Q | 2.14 | {15} |
| 12. | R->G | 8.32 | {4} | | G->R | 2.08 | {4} |
| 13. | G->D | 8.0 | {3} | 22. | B->H | 1.92 | {7} |
| | G->N | 5.82 | {11} | 23. | Q->U | 1.89 | {18} |
| | S->G | 5.78 | {8} | | N->K | 1.52 | {14} |
| | G->A | 5.78 | {1} | | M->I | 1.45 | {22} |
| | F->G | 5.2 | {9} | | H->B | 1.45 | {7} |
| 14. | I->P | 5.19 | {16, 20} | | M->L | 1.3 | {24, 27} |
| | P->I | 5.09 | {20, 27} | | O->N | 1.21 | {12} |
| | G->P | 4.90 | {27} | 24. | N->T | 1.21 | {13} |
| 15. | E->O | 4.40 | {21, 25} | | | | |

Table 3 shows that 24 dependencies are removed to make CDG acyclic. The edges highlighted bold in Table 3 are neglected, as the cycles in which they are occurring have already been deleted. Resulting acyclic CDG is shown in Figure 5.

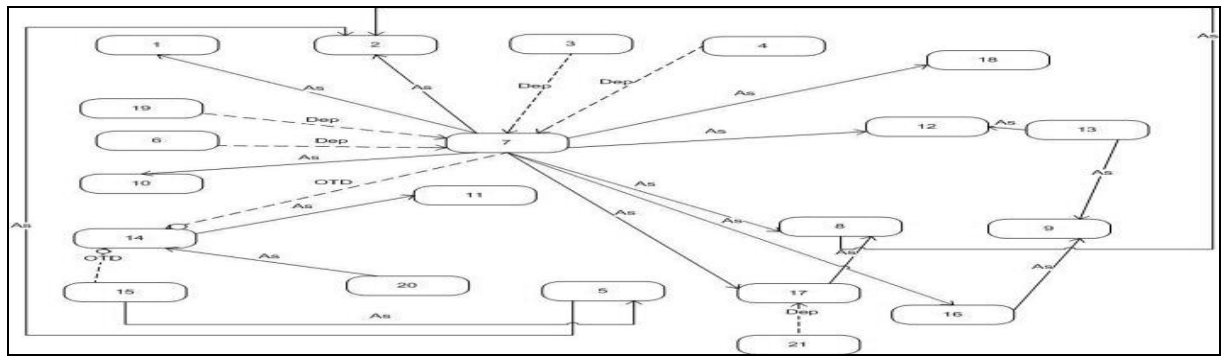

**Figure 5.** An Acyclic CDG of ATM Class Diagram

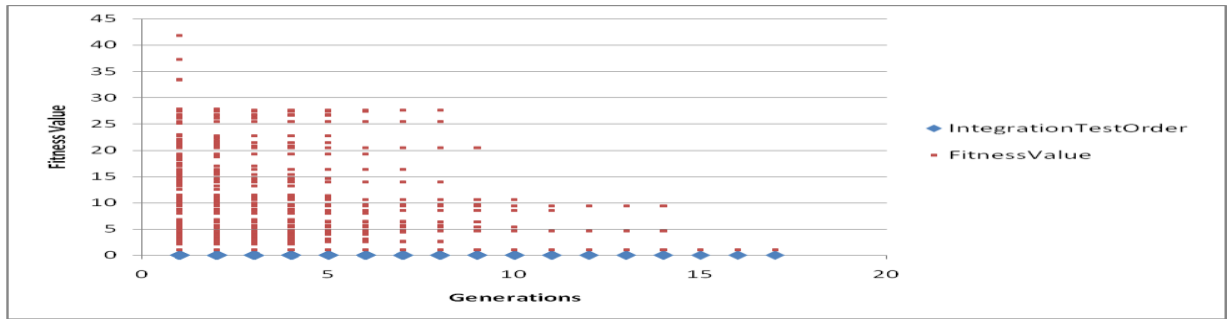

**Figure 6.** Output of ATM Class Diagram

The experiment is performed on .Net framework using C sharp. In order to apply GA, we start with an initial population of randomly generated test orders or individuals. For the test order involving number of classes {A, B, C, D, E, F, G, H, I, J, K, L, M, N, O, P, Q, R, S, T and U}, a possible chromosome sequence is {B, J, K, S, L, C, M, A, O, N, P, D, Q, G, I, U, H, T, E, S, F, R}. One point crossover is applied to generate new individuals. The GA is run for 20 generations with an initial population of 100 individuals.

The specific stubs in CDG of ATM system are L->G, A->G, G->K, H->E, E->H, H->I, G->F, N->G, Q->S, G->S, J->G, R->G, G->D, I->P, E->O, G->C, B->E, H->Q, I->M, K->N, N->O, B->H and Q->U and N->T. The realistic stubs are C, D, E, F, G, H, I, K, M, N, O, P, Q, S, T, and U. The class C is tested using stub G, class D is tested using stub G, class E is tested using stub B and stub H, class F is tested using stub G, class G is tested using stubs A, J, N, Q and R, class H is tested using stubs E and stub B. The specific stubs are twenty four and realistic stubs are sixteen in ATM case study. By applying GA, results have shown that the class integration order GOKAQRNSCDJLFHTPUIEBM has lowest weight of 1.1 among other individuals in the generations.

We applied topological sorting to derive integration test order. When a choice between vertices is made, we use vertices ordering. For example, when a choice between vertices C and D is made, we choose C. By applying topological ordering, the integration order for whole CDG is {G < A, J, N, Q, R >,O < E, N >, K < G >, A , Q < H >, R, N < K >, S < Q , G >, C  < G >, D < G > , J, L, F < G >, H < E, B >, T < N >, P < I >, U < Q >, I < H >, E < H, B >, B, M < I >}.The real components to test stubs are shown in bracket < >. The number of integration steps is 45. The number of steps is quite large because of cyclic CDG. The integration cost of ATM system is 167.23, calculated by adding the weight of the edges removed in order to make the CDG acyclic. The results of our ATM case study are shown in the Figure 6. The graph plots the weight or fitness value of individuals in each generation. As the number of generation increases, the fitness value of individuals starts decreasing.

## 6. Comparison of the Proposed Approach with Existing Techniques

In this section we perform a comparison of our proposed approach for solving CITO problem with existing graph based and GA based techniques.

## 6.1 Comparison of Proposed Approach with Existing Graph based Approach

In Table 3, we have shown number of edges removed for converting cyclic CDG of ATM into acyclic CDG. In this section, we are comparing graph based heuristic by Abduzarik's [12] for breaking cycles with our algorithm. The algorithm for breaking cycles in WORD by Abduzarik's strategy is shown in Figure 7.

```
1:  Find all SCCs in WORD
2:  for (each scc_i(V_scc_i, E_scc_i) ∈ SCCs) do
3:    find all cycles CYCLES (totalCycles)
4:    for (each e ∈ E_scc_i) do
5:      find the number of cycles that use e
          (cardinal{cycles − through − e})
6:      compute the cycle-weight ratio
7:    end for
8:    while (totalCycles != 0) do
9:      order all edges in descending order of their
        cycle-weight ratio
10:     remove edge with highest cycle-weight ratio
11:     totalCycle = totalCycle - number of cycles
        broken
12:     update the number of cycles that use e
        (cardinal{cycles − through − e}) in the
        remaining edge set
13:     recompute the cycle-weight ratio for the
        remaining edges
14:   end while
15: end for
```

**Figure 7.** Abduzarik's Algorithm for eliminating cycles in WORD [12]

I. Table 4 shows the steps of breaking cycles by applying Aduzarik's strategy. Remove cycle H->E having CWR 25 and update CWR for remaining edges.

**Table 4.** Steps for converting Cyclic CDG into Acyclic CDG

| S. No. | Edge | CS (Weight) | Cycles Removed | Number of cycles (NC) | Cycle to weight ratio (CWR) | S. No. | Edge | CS (Weight) | Cycles Removed | Number of cycles (NC) | Cycle to weight ratio (CWR) |
|---|---|---|---|---|---|---|---|---|---|---|---|
| 1. | A->G | 0.09 | {1} | 1 | 11.11 | 25. | E->B | 0.4 | {23} | 1 | 2.5 |
| 2. | G->A | 0.18 | {1} | 1 | 5.56 |  | H->E | 0.08 | {6, 25} | 2 | 25 |
| 3. | F->G | 0.2 | {9} | 1 | 5.0 | 26. | E->H | 0.08 | {26} | 1 | 12.5 |
| 4. | G->F | 0.1 | {9} | 1 | 10 | 27. | G->H | 0.46 | {16, 17, 24} | 3 | 6.52 |
| 5. | J->G | 0.13 | {10} | 1 | 7.69 | 28. | H->I | 0.08 | {16, 24} | 2 | 25 |
| 6. | G->J | 0.25 | {10} | 1 | 4.0 | 29. | H->Q | 0.3 | {17, 28} | 2 | 6.67 |
| 7. | C->G | 0.38 | {2} | 1 | 2.63 | 30. | Q->H | 0.2 | {28} | 1 | 5 |
| 8. | G->C | 0.25 | {2} | 1 | 4.0 | 31. | G->Q | 0.56 | {15} | 1 | 1.79 |
| 9. | D->G | 0.25 | {3} | 1 | 4.0 | 32. | Q->G | 0.22 | {15, 17} | 2 | 9.09 |
| 10. | G->D | 0.13 | {3} | 1 | 7.69 | 33. | I->P | 0.11 | {16, 20} | 2 | 18.18 |
| 11. | S->G | 0.18 | {8} | 1 | 5.56 | 34. | P->I | 0.11 | {20, 27} | 2 | `18.18 |
| 12. | G->S | 0.11 | {8} | 1 | 9.09 | 35. | G->P | 0.22 | {27} | 1 | 4.55 |
| 13. | R->G | 0.13 | {4} | 1 | 7.69 | 36. | I->M | 0.33 | {22, 24, 27} | 3 | 9.09 |
| 14. | G->R | 0.5 | {4} | 1 | 2 | 37. | M->I | 0.22 | {22} | 1 | 4.55 |
| 15. | L->G | 0.22 | {5, 24, 27} | 3 | 13.64 | 38. | M->L | 0.33 | {24, 27} | 2 | 6.06 |
| 16. | G->L | 0.33 | {5} | 1 | 3.03 | 39. | G->K | 0.1 | {19} | 1 | 10 |
| 17. | U->Q | 0.33 | {18} | 1 | 3.03 | 40. | K->N | 0.29 | {14, 19} | 2 | 0.07 |
| 18. | Q->U | 0.17 | {18} | 1 | 5.88 | 41. | N->K | 0.29 | {11} | 1 | 3.45 |
| 19. | N->G | 0.33 | {11, 19} | 2 | 6.06 | 42. | N->T | 0.33 | {13} | 1 | 3.03 |
| 20. | G->N | 0.22 | {11} | 1 | 4.55 | 43. | T->N | 0.33 | {13} | 1 | 3.03 |
| 21. | N->O | 0.17 | {12} | 1 | 5.88 | 44. | E->O | 0.13 | {21} | 1 | 7.69 |
| 22. | O->N | 0.33 | {12} | 1 | 3.03 | 45. | O->E | 0.38 | {21} | 1 | 2.63 |
| 23. | G->B | 0.27 | {26} | 1 | 3.70 | 46. | B->H | 0.25 | {7} | 1 | 4.0 |
| 24. | B->E | 0.2 | {23, 26} | 2 | 10 | 47. | H->B | 0.33 | {7} | 1 | 3.03 |

II. Remove edge H->I having CWR 25 and update CWR for remaining edges. The Table 5 shows this process.

**Table 5.** Steps for converting Cyclic CDG into Acyclic CDG after removing Edge H->E

| S. No. | Edge | CS (Weight) | Cycles Removed | Number of cycles (NC) | Cycle to weight ratio (CWR) | S. No. | Edge | CS (Weight) | Cycles Removed | Number of cycles (NC) | Cycle to weight ratio (CWR) |
|---|---|---|---|---|---|---|---|---|---|---|---|
| 1. | A->G | 0.09 | {1} | 1 | 11.11 | 25. | E->B | 0.4 | {23} | 1 | 2.5 |
| 2. | G->A | 0.18 | {1} | 1 | 5.56 | | H->E | 0.08 | {6, 25} | 2 | 25 |
| 3. | F->G | 0.2 | {9} | 1 | 5.0 | 26. | E->H | 0.08 | {26} | 1 | 12.5 |
| 4. | G->F | 0.1 | {9} | 1 | 10 | 27. | G->H | 0.46 | {17} | 1 | 2.17 |
| 5. | J->G | 0.13 | {10} | 1 | 7.69 | | H->I | 0.08 | {16, 24} | 2 | 25 |
| 6. | G->J | 0.25 | {10} | 1 | 4.0 | 28. | H->Q | 0.3 | {17, 28} | 2 | 6.67 |
| 7. | C->G | 0.38 | {2} | 1 | 2.63 | 29. | Q->H | 0.2 | {28} | 1 | 5 |
| 8. | G->C | 0.25 | {2} | 1 | 4.0 | 30. | G->Q | 0.56 | {15} | 1 | 1.79 |
| 9. | D->G | 0.25 | {3} | 1 | 4.0 | 31. | Q->G | 0.22 | {15, 17} | 2 | 9.09 |
| 10. | G->D | 0.13 | {3} | 1 | 7.69 | 32. | I->P | 0.11 | {20} | 1 | 9.09 |
| 11. | S->G | 0.18 | {8} | 1 | 5.56 | 33. | P->I | 0.11 | {20, 27} | 2 | `18.18 |
| 12. | G->S | 0.11 | {8} | 1 | 9.09 | 34. | G->P | 0.22 | {27} | 1 | 4.55 |
| 13. | R->G | 0.13 | {4} | 1 | 7.69 | 35. | I->M | 0.33 | {22, 27} | 2 | 6.06 |
| 14. | G->R | 0.5 | {4} | 1 | 2 | 36. | M->I | 0.22 | {22} | 1 | 4.55 |
| 15. | L->G | 0.22 | {5, 27} | 2 | 9.09 | 37. | M->L | 0.33 | {27} | 1 | 3.03 |
| 16. | G->L | 0.33 | {5} | 1 | 3.03 | 38. | G->K | 0.1 | {19} | 1 | 10 |
| 17. | U->Q | 0.33 | {18} | 1 | 3.03 | 39. | K->N | 0.29 | {14, 19} | 2 | 0.07 |
| 18. | Q->U | 0.17 | {18} | 1 | 5.88 | 40. | N->K | 0.29 | {11} | 1 | 3.45 |
| 19. | N->G | 0.33 | {11, 19} | 2 | 6.06 | 41. | N->T | 0.33 | {13} | 1 | 3.03 |
| 20. | G->N | 0.22 | {11} | 1 | 4.55 | 42. | T->N | 0.33 | {13} | 1 | 3.03 |
| 21. | N->O | 0.17 | {12} | 1 | 5.88 | 43. | E->O | 0.13 | {21} | 1 | 7.69 |
| 22. | O->N | 0.33 | {12} | 1 | 3.03 | 44. | O->E | 0.38 | {21} | 1 | 2.63 |
| 23. | G->B | 0.27 | {26} | 1 | 3.70 | 45. | B->H | 0.25 | {7} | 1 | 4.0 |
| 24. | B->E | 0.2 | {23, 26} | 2 | 10 | 46. | H->B | 0.33 | {7} | 1 | 3.03 |

III. For space reasons, all steps for breaking cycles are not shown. The order of dependencies deleted after deleting H->E are P->I, E->H, A->G, G->F, G->K, Q->G, G->S, J->G, G->D, R->G, E->O, Q->U, N->O, Q->H, B->E, L->G, G->N, M->I, G->C, B->H, K->N and N->T. Hence, 24 dependencies are removed to make graph acyclic.

The total number of dependencies removed from our proposed and graph based strategy [12] are same. In our approach, dependencies removed to make CDG acyclic are different. In our proposed approach, less computation effort is required as the edges are removed without recomputing weights for all edges in CDG.

### 6.2 Comparison of Proposed Approach with existing GA based Approach
To compare our algorithm with GA based approach, we have applied our algorithm on the Briand's ATM case study. We have used the dependency tables [13] as shown in Figure 8.

| | 1 | 2 | 3 | 4 | 5 | 6 | 7 | 8 | 9 | 10 | 11 | 12 | 13 | 14 | 15 | 16 | 17 | 18 | 19 | 20 | 21 |
|---|---|---|---|---|---|---|---|---|---|---|---|---|---|---|---|---|---|---|---|---|---|
| 1 | | | | | | | | | | | | | | | | Us | | | | Us | |
| 2 | | | | | | | | | | | | | | | | Us | | | | | |
| 3 | | Us | | | | | | | | | | | | | | | | | | Us | |
| 4 | | | | | | | | | | | | | | | | | Us | | | Us | |
| 5 | | | | | | | | | | | | | | | | | Us | | | Us | |
| 6 | | | | | | | | | | | | | | | | | | | | | |
| 7 | | | | | | | | | | | | | | | | | | | | Cp | |
| 8 | Cp | Cp | Cp | Cp | Cp | Cp | Cp | | As | Us | | | | | | | Us | | | Us | |
| 9 | | | | | | | | Us | | | | | | | | | | | | Cp | Us |
| 10 | | | | | | | | As | As | | Cp | Us | Us | Us | Us | | | | | | Us |
| 11 | | | | | | | | As | As | As | | | | | | | | | | Cp | Us |
| 12 | | | | | | | | Us | Us | Us | I | | | | | | | | | As | Us |
| 13 | | | | | | | | Us | Us | Us | I | | | | | | | | | As | Us |
| 14 | | | | | | | | Us | Us | Us | I | | | | | | | | | As | Us |
| 15 | | | | | | | | Us | Us | Us | I | | | | | | | | | Us | Us |
| 16 | | | | | | | | | | | | | | | | | | | | | |
| 17 | | | | | | | | | | | | | | | | Us | | | | | |
| 18 | | | | | | | | Cp | Cp | | | | | | | | | | | Us | |
| 19 | | | | | | | | Cp | Cp | | | | | | | | | | | Us | |
| 20 | | | | | | | | | | | | | | | | | | | | | |
| 21 | | | | | | | | | | | | | | | | | | | | | |

**Figure 8.** ATM Dependency Matrix. As = Association, Cp = Composition, Us = Use dependency, I = Inheritance [13]

We have transformed dependency table shown in Figure 8 into CDG and identified thirty cycles which are as follows:-

1. 8->9->8
2. 8->10->8
3. 10->9->8->10
4. 10->11->8->10
5. 10->11->9->8->10
6. 10->12->8->10
7. 10->13->8->10
8. 10->14->10
9. 10->14->8->10
10. 10->15->8->10
11. 10->12->9->8->10
12. 10->13->9->8->10
13. 10->14->9->8->10
14. 10->15->9->8->10
15. 10->13->10
16. 10->15->10
17. 12->11->10->12
18. 13->11->10->13
19. 13->11->8->10->13
20. 13->11->9->8->10->13
21. 14->11->8->10->14
22. 14->11->9->8->10->14
23. 14->11->10->14
24. 15->11->9->8->10->15
25. 15->11->10->15
26. 15->9->8->10->15
27. 10->11->10
28. 10->12->10
29. 15->11->8->10->15
30. 12->11->8->10->12

**Table 6.** Coupling measures for SCC (8, 9, 10, 11, 12, 13, 14, 15) [12]

| No. | 8 | 9 | 10 | 11 | 12 | 13 | 14 | 15 |
|---|---|---|---|---|---|---|---|---|
| 8 | | 1.0.0.0.0 | 1.0.2.1.3 | | | | | |
| 9 | 1.0.1.1.3 | | | | | | | |
| 10 | 1.0.6.4.4 | 1.0.1.1.1 | | 1.0.3.5.3 | 1.0.1.1.0 | 10.1.1.0 | 1.0.1.1.0 | 1.0.1.1.0 |
| 11 | 1.0.1.1.3 | 1.0.0.0.0 | 1.0.1.1.0 | | | | | |
| 12 | 1.0.4.3.11 | 1.0.4.3.11 | 1.0.2.2.0 | 5.0.0.0.0 | | | | |
| 13 | 1.0.4.3.6 | 1.0.4.3.14 | 1.0.2.2.0 | 5.0.0.0.0 | | | | |
| 14 | 1.0.3.2.6 | 1.0.3.3.12 | 1.0.2.2.0 | 5.0.0.0.0 | | | | |
| 15 | 1.0.2.1.6 | 1.0.3.3.10 | 1.0.2.2.0 | 5.0.0.0.0 | | | | |

**Table 7.** Edge Weights for SCC { 8, 9, 10, 11, 12, 13, 14, 15}

| Edge | D (1) | # of Attr. (2) | # of Meth. (3) | A & M (4) | A&M [12] (5) | CS (6) | CW | IF | $W_i$ (1) | $W_i$ (2) | $W_i$ (3) | $W_i$ (4) | $W_i$ (5) | $W_i$ (6) |
|---|---|---|---|---|---|---|---|---|---|---|---|---|---|---|
| 8->9 | 1 | 13 | 1 | 0.71 | 1 | 1 | 0.03 | 9 | 0.27 | 0.02 | 0.27 | 0.38 | 0.27 | 0.27 |
| 8->10 | 1 | 9 | 2 | 0.53 | 1.22 | 7 | 0.7 | 20 | **14** | **1.56** | **7** | **26.41** | **11.48** | **2** |

| | | | | | | | | | | | | |
|---|---|---|---|---|---|---|---|---|---|---|---|---|
| **9->8** | 1 | 13 | 7 | 1 | 1.17 | 6 | 0.4 | 17 | **6.8** | **0.52** | **0.97** | **6.8** | **5.81** | **1.13** |
| 10->8 | 1 | 13 | 7 | 1 | 1.66 | 15 | 0.03 | 22 | 0.66 | 0.05 | 0.09 | 0.66 | 0.40 | 0.044 |
| 10->9 | 1 | 13 | 2 | 0.74 | 1.13 | 4 | 0.03 | 21 | 0.63 | 0.05 | 0.315 | 0.85 | 0.56 | 0.16 |
| 10>11 | 1 | 0 | 0 | 0 | 1.57 | 12 | 0.1 | 21 | 2.1 | - | - | - | 1.34 | 0.18 |
| **10->12** | 1 | 2 | 2 | 0.23 | 1.13 | 3 | 0.17 | 18 | **3.06** | **1.53** | **1.53** | **13.30** | **2.71** | **1.02** |
| **10->13** | 1 | 2 | 2 | 0.23 | 1.13 | 3 | 0.2 | 18 | **3.6** | **1.8** | **1.8** | **15.65** | **3.19** | **1.2** |
| **10->14** | 1 | 3 | 2 | 0.26 | 1.13 | 3 | 0.2 | 18 | **3.6** | **1.2** | **1.8** | **13.85** | **3.19** | **1.2** |
| **10->15** | 1 | 1 | 2 | 0.21 | 1.13 | 3 | 0.23 | 18 | **4.14** | **4.14** | **2.07** | **19.71** | **3.66** | **1.38** |
| 11->8 | 1 | 13 | 2 | 0.74 | 1.17 | 6 | 0.13 | 17 | 2.21 | 0.17 | 1.11 | 2.99 | 1.89 | 0.37 |
| 11->9 | 1 | 13 | 1 | 0.71 | 1.00 | 1 | 0.13 | 16 | 2.08 | 0.21 | **2.72** | 3.83 | **2.72** | **2.08** |
| **11->10** | 1 | 9 | 2 | 0.53 | 1.13 | 3 | 0.13 | 21 | **2.73** | **0.30** | **1.37** | **5.15** | **2.42** | **0.91** |
| 12->8 | 1 | 13 | 4 | 0.81 | 1.60 | 19 | 0.03 | 14 | 0.42 | 0.03 | 0.11 | 0.52 | 0.26 | 0.02 |
| 12->9 | 1 | 13 | 4 | 0.81 | 2.59 | 19 | 0.03 | 13 | 0.39 | 0.03 | 0.10 | 0.48 | 0.15 | 0.02 |
| 12->10 | 1 | 9 | 2 | 0.53 | 2.01 | 5 | 0.03 | 18 | 0.54 | 0.06 | Inf | 1.01 | 0.27 | 0.11 |
| 12->11 | 1 | Inf | Inf | Inf | 5.00 | 5 | 0.03 | 13 | 0.39 | Inf | Inf | Inf | 0.08 | 0.08 |
| 13->8 | 1 | 13 | 4 | 0.77 | 1.50 | 14 | 0.03 | 14 | 0.42 | 0.03 | 0.11 | 0.52 | 0.28 | 0.03 |
| 13->9 | 1 | 13 | 4 | 0.77 | 2.62 | 22 | 0.03 | 13 | 0.39 | 0.03 | 0.10 | 0.48 | 0.19 | 0.02 |
| 13->10 | 1 | 9 | 2 | 0.53 | 2.01 | 5 | 0.03 | 18 | 0.54 | 0.06 | 0.27 | 1.07 | 0.27 | 0.11 |
| 13->11 | 1 | Inf | Inf | Inf | 5.00 | 5 | 0.1 | 13 | 1.3 | Inf | Inf | Inf | 0.26 | 0.26 |
| 14->8 | 1 | 13 | 3 | 0.74 | 1.39 | 12 | 0.03 | 14 | 0.42 | 0.03 | 0.14 | 0.55 | 0.30 | 0.04 |
| 14->9 | 1 | 13 | 3 | 0.77 | 2.58 | 19 | 0.03 | 13 | 0.39 | 0.03 | 0.13 | 0.51 | 0.15 | 0.02 |
| 14->10 | 1 | 9 | 2 | 0.43 | 2.01 | 5 | 0.03 | 18 | 0.54 | 0.06 | 0.27 | 1.01 | 0.27 | 0.11 |
| 14->11 | 1 | Inf | Inf | Inf | 5.00 | 5 | 0.1 | 13 | 1.3 | Inf | Inf | Inf | 0.26 | 0.26 |
| 15->8 | 1 | 13 | 2 | 0.74 | 1.29 | 10 | 0.03 | 14 | 0.42 | 0.03 | 0.21 | 0.57 | 0.33 | 0.04 |
| 15->9 | 1 | 13 | 3 | 0.77 | 2.58 | 17 | 0.03 | 13 | 0.39 | 0.03 | 0.13 | 0.51 | 0.15 | 0.02 |
| 15->10 | 1 | 9 | 2 | 0.53 | 2.01 | 5 | 0.03 | 18 | 0.54 | 0.06 | 0.27 | 1.01 | 0.27 | 0.11 |
| 15->11 | 1 | Inf | Inf | Inf | 5.00 | 5 | 0.1 | 13 | 1.3 | Inf | Inf | Inf | 0.26 | 0.26 |

Applying Tarjan's algorithm, SCC identified is {8, 9, 10, 11, 12, 13, 14, and 15}. In Briand's work, edge weight is obtained from the four cost functions which are attributes coupling, method coupling, number of broken dependencies and weighted geometric average of methods. Since the number of attributes and methods in Briand's ATM case study were not available to us, we have taken edge weights same as provided in their work. We extracted the data from Abduzarik's et. al. work [12] to determine Briand's number of broken dependencies (D (1)), attribute coupling (# of Attr (2)), method coupling (# of Method.(3)), attributes and method coupling (A & M (4)) and attributes and method coupling (A & M (5)) as shown in Table 7.

Using dependency table shown in Figure 8, we calculated IF complexity for each class in ATM case study by applying equation 3. We calculated CS between two classes $c_i$ and $c_j$ for $W_i$ (6) by adding five parameters shown in Table 6 i.e. C. $V_d$. $M_d$. $R_d$. $P_d$ where, C is 5 for inheritance and composition dependency and 1 for other dependencies, $V_d$ is number of distinct public variables of $c_j$ directly used by $c_i$, $M_d$ is number of public methods of $c_j$ that are called by $c_i$, $R_d$ is number of distinct return typed that occur in $M_d$, $P_d$ is number of distinct parameters that occur in $M_d$ [12]. CW is calculated in the manner as explained in section 5. The number of broken dependencies (1), attribute coupling (2), method coupling (3), attribute and method coupling (4) and attribute and method coupling (5) are used to determine CS in our work. Applying equation 4, we have calculated $W_i$ for six cost functions as shown in Table 7. The results shows that for each cost function, 7 dependencies are removed (8->10, 9->8, 10->12, 10->13, 10->14, 10->15 and 11->10).

The stub cost for six cost functions are equal as the same set of dependencies were removed. In section 6.1 and section 6.2, our illustrative case studies show that parameters i.e. CS, CW and IF as depicted in our work are effective in determining the test order of minimum stub complexity.

## 7. Comparison of GA based Proposed Approach with Micro-GA and Cuckoo Search Algorithm

Table 8. Parameters for GA, Micro-GA and Cuckoo Search Algorithm

| Parameters | GA | Micro-GA | Cuckoo Search |
|---|---|---|---|
| Initial Population Size | 100 | 5 | 100 |
| Selection | Average weight = 12 | Elitism | If Fitness $_{new\ individual}$ > Fitness $_{old\ individual}$. Replace old individual with new individual. |
| Crossover Probability | Weight > Fitness $_{Average}$, Crossover Probability = 75%  Weight < Fitness $_{Average}$, Crossover Probability = 25% | 100% for 4 leftover individuals. | ----- |
| Mutation Probability | Weight > Fitness $_{Average}$, Mutation Probability = 25%  Weight < Fitness $_{Average}$, Mutation Probability = 15% | ----- | Delete 20 worst individuals in each generation. |

Table 9. Fitness value of Individuals

| Micro-GA | | | Cuckoo Search Algorithm | | | GA | | |
|---|---|---|---|---|---|---|---|---|
| Gen | Average Weight | Minimum Weight | Gen | Average Weight | Minimum Weight | Gen | Average Weight | Minimum Weight |
| 1 | 15.794 | 6.03 | 1 | 12.2032 | 0.97 | 1 | 13.9159 | 1.1 |
| 2 | 10.288 | 5.52 | 2 | 11.8211 | 0.97 | 2 | 12.56237 | 1.1 |
| 3 | 8.906 | 3.96 | 3 | 10.8078 | 0.97 | 3 | 11.58508 | 1.1 |
| 4 | 8.88 | 3.96 | 4 | 10.5146 | 0.97 | 4 | 12.18 | 1.1 |
| 5 | 14.308 | 2.08 | 5 | 9.7183 | 0.97 | 5 | 12.174 | 1.1 |
| 6 | 11.364 | 2.08 | 6 | 9.4754 | 0.97 | 6 | 11.29393 | 1.1 |
| 7 | 11.784 | 2.08 | 7 | 9.5483 | 0.97 | 7 | 10.01045 | 1.1 |
| 8 | 5.332 | 2.08 | 8 | 9.2575 | 0.97 | 8 | 10.03789 | 1.1 |
| 9 | 9.892 | 1.45 | 9 | 9.7958 | 0.97 | 9 | 8.573077 | 1.1 |
| 10 | 12.422 | 1.45 | 10 | 9.2989 | 0.97 | 10 | 6.78 | 1.1 |
| 11 | 6.698 | 1.45 | 11 | 9.1869 | 0.97 | 11 | 4.978 | 1.1 |
| 12 | 10.644 | 1.45 | 12 | 8.9466 | 0.97 | 12 | 5.076667 | 1.1 |
| 13 | 12.224 | 1.45 | 13 | 9.0851 | 0.97 | 13 | 5.076667 | 1.1 |
| 14 | 10.724 | 1.45 | 14 | 9.5793 | 0.97 | 14 | 5.076667 | 1.1 |
| 15 | 7.662 | 1.45 | 15 | 9.424 | 0.97 | 15 | 1.1 | 1.1 |
| 16 | 9.024 | 0.84 | 16 | 9.3312 | 0.97 | 16 | 1.1 | 1.1 |
| 17 | 8.982 | 0.84 | 17 | 8.8746 | 0.97 | 17 | 1.1 | 1.1 |
| 18 | 7.928 | 0.84 | 18 | 9.0394 | 0.97 | 18 | 1.1 | 1.1 |
| 19 | 6.44 | 0.84 | 19 | 8.624 | 0.97 | 19 | 1.1 | 1.1 |
| 20 | 5.482 | 0.84 | 20 | 8.6242 | 0.97 | 20 | 1.1 | 1.1 |
| 21 | 11.812 | 0.84 | 21 | 8.1939 | 0.97 | 21 | 1.1 | 1.1 |
| 22 | 10.664 | 0.84 | 22 | 8.1224 | 0.97 | 22 | 1.1 | 1.1 |

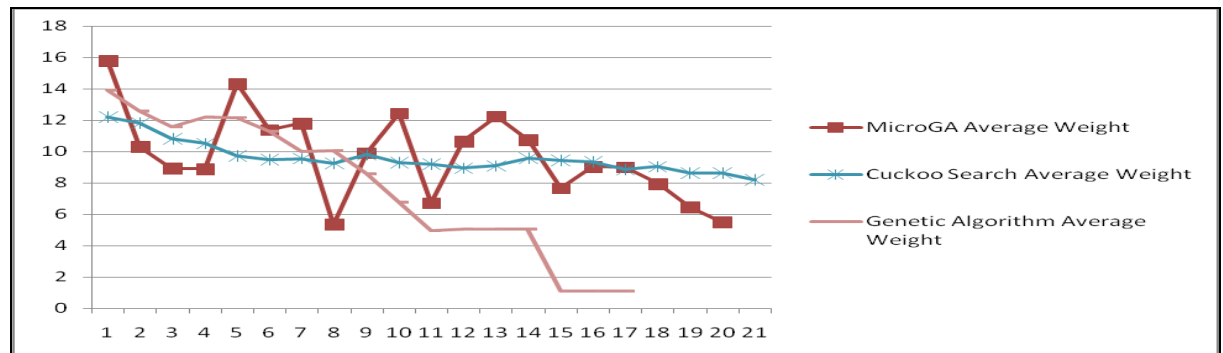

**Figure 9.** GA, Micro-GA, and Cuckoo Search Algorithm results on Proposed Approach

We have applied GA, micro-GA and cuckoo search algorithm on ATM case study provided in section 5. Table 8 shows different parameters values for GA, micro-GA and cuckoo search algorithm. The average fitness value of individuals after applying GA, micro- GA and cuckoo search algorithm are shown in Table 9. The average fitness value is determined by adding total fitness value of each individual in current generation divided by total number of individuals. The graphical results of applying GA, micro- GA and cuckoo search algorithm is shown in Figure 9 where X-axis plots the number of generations and Y-axis plots the average fitness value. The average fitness value decreases sharply by applying micro-GA but as there are a majority of new values in every generation; the results vary and do not decrease with each generation. On applying cuckoo search algorithm, the average fitness decreases but the results shows that initially, micro-GA outperforms GA and cuckoo search algorithm. As the run time increases, GA gives better results compared to micro-GA and cuckoo search algorithm.

For large case studies, GA gives better results than micro-GA and cuckoo search algorithm. It can be concluded that micro-GA and cuckoo search algorithm gives better results for the small sized problems or when the time constraint is limited.

## 8. Conclusion and Future Work

In this paper an approach is proposed to identify the test order having minimum cost by applying GA. CS is used to capture intensity of interaction in terms of number of methods and attributes coupled between client and server class. We have also compared the results of GA, micro-GA and cuckoo search algorithm on our ATM case study. The empirical result shows that GA outperforms micro-GA and cuckoo search algorithm for large sized problems. The proposed approach is effective for solving problems where test order size is very large. The results of our proposed approach are as good as the results produced by existing graph based and other GA based techniques, but the computation effort is less with our proposed approach as compared to existing graph based approaches.

Limitation:-
- The test dependencies for realization, binding and extend relationship in UML class diagram has not been investigated in our work.
- The proposed GA based approach is effective for large sized group of classes.

Our future work involves proposing a new cost effective technique using GA and graph based approach. In future, we plan to analyze the test dependencies for realization, binding and extend relationship .We will also compare our approach to other existing approaches. A tool is also being developed to support this proposed approach.